\documentclass[twocolumns]{sdl2}

\def\0{\mbox{\tiny $0$}}
\def\1{\mbox{\tiny $1$}}
\def\2{\mbox{\tiny $2$}}
\def\3{\mbox{\tiny $3$}}
\def\4{\mbox{\tiny $4$}}
\def\5{\mbox{\tiny $5$}}
\def\6{\mbox{\tiny $6$}}
\def\7{\mbox{\tiny $7$}}
\def\8{\mbox{\tiny $8$}}
\def\9{\mbox{\tiny $9$}}

\definecolor{C1}{rgb}{0.18,0.18,1}
\definecolor{C2}{rgb}{0.8,0.35,0.0}
\definecolor{C3}{rgb}{0.0,0.5,0.0}
\def\CC{White}
\def\CS{White}

\def\wo{\mathrm{w}_{\0}}
\def\wos{\mathrm{w}_{\0}^{\2}}
\def\w{\mathrm{w}}
\def\d{\mathrm{d}}
\def\zR{z_{_{\mathrm{R}}}}
\def\q{^{^2}}
\def\qq{^{\2}}

\def\Gtra{G\pol\tr(\kx,\ky)}

\def\in{^{^{\mathrm{[inc]}}}}

\def\tr{^{^{\mathrm{[tra]}}}}
\def\pol{_{_{\mathrm{pol}}}}
\def\tm{_{_{\mathrm{tm}}}}
\def\te{_{_{\mathrm{te}}}}

\def\sn{_{_{\mathrm{Snell}}}}

\def\xt{\widetilde{x}}

\def\zt{\widetilde{z}}

\def\xs{x_*}

\def\zs{z_*}

\def\rtr{\mathbf{r}_{_{\mathrm{tra}}}}
\def\rttr{\widetilde{r}_{_{\mathrm{tra}}}}
\def\xttr{\widetilde{x}_{_{\,\mathrm{tra}}}}
\def\xtr{x_{_\mathrm{tra}}}

\def\ztr{z_{_\mathrm{tra}}}

\def\mk{|\boldsymbol{k}|}
\def\kx{k_{_x}}
\def\ky{k_{_y}}
\def\kz{k_{_z}}

\def\kxt{k_{\tilde{x}}}

\def\kzt{k_{\tilde{z}}}
\def\qzt{q_{\tilde{z}}}
\def\qxt{q_{\tilde{x}}}

\def\kzs{k_{z_{*}}}
\def\qzs{q_{z_{*}}}
\def\qxs{q_{x_{*}}}

\def\max{^{^{\mathrm{[max]}}}}

\def\ua{^{^{[1]}}}
\def\ub{^{^{[2]}}}
\def\uc{^{^{[3]}}}
\def\uac{^{^{[1+3]}}}

\newcommand*{\circledM}[3][]{\tikz[baseline=(C.base)]{
	\node[draw, circle, inner sep=1pt, yshift=1pt, fill=#3](C) 
        {\vphantom{1g}};
    \node[inner sep=0pt] at (C.center)
        {\color{White}$\mathbf{#2}$};
    }}
    
    \newcommand*{\circledO}[3][]{\tikz[baseline=(C.base)]{
	\node[draw, circle, inner sep=1pt, yshift=1pt, fill=#3](C) 
        {\vphantom{1g}};
    \node[inner sep=0pt] at (C.center)
        {\color{Black}$\mathbf{#2}$};
    }}
\newcommand{\refa}[6]{{#1,}{ \em #2,}{ #3}{ \textbf{#4}}{, #5}{ (#6).}} %\refa{author}{title}{journal}{vol}%{page}{year}
\newcommand{\refb}[4]{{#1,}{ #2}{ (#3,}{ #4).}} %  \referbook{author}{title}{publisher}{year}

\def\figureone{
\WideFigureSideCaption{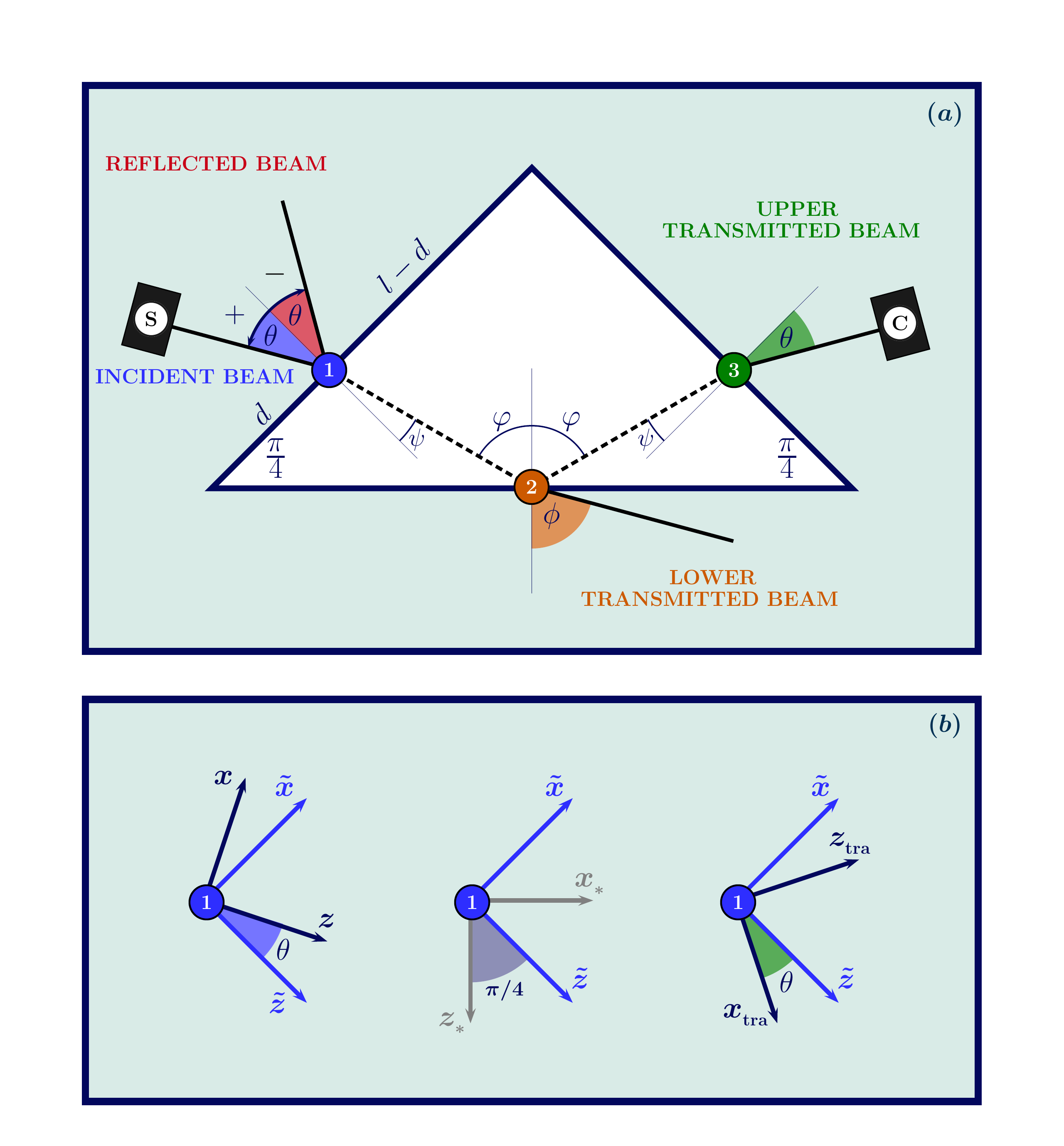}{{\small \textbf{Geometrical layout of the dielectric prism.}} In (a), a laser beam moves, along the $z$-axis, from the source, \circledO{S}{\CS}, to the air/dielectric interface, \circledM{1}{C1},   forming an incidence angle,  $\theta$, with the normal to the first interface, $\zt$. The  beam transmitted through the first interface then moves in the dielectric prism towards  the second  (dielectric/air) interface, \circledM{2}{C2}, forming an angle $\psi$ with the normal to the first interface, $\zt$, and an angle $\varphi$ with the normal to the second interface, 
$\zs$. These angles are related to the incidence one by the Snell law, $\sin\theta=n\,\sin\psi$ and $n\,\sin\varphi=\sin\phi$, where  $\varphi=\psi+\pi/4$.  Once reflected by the second interface, the optical beam moves to the third (dielectric/air) interface,      \circledM{3}{C3}. Due to the geometry of the prism, the upper transmitted beam 
forms an angle $\theta$ with respect to the normal to the third interface, $\xt$.
The upper transmitted beam is thus detected at the camera \circledO{C}{\CC}. In (b), we find  the coordinates systems of the incident and  upper transmitted beams, and  of the prism interfaces. \label{fig1}}}

\def\figuretwo{
\WideFigureSideCaption{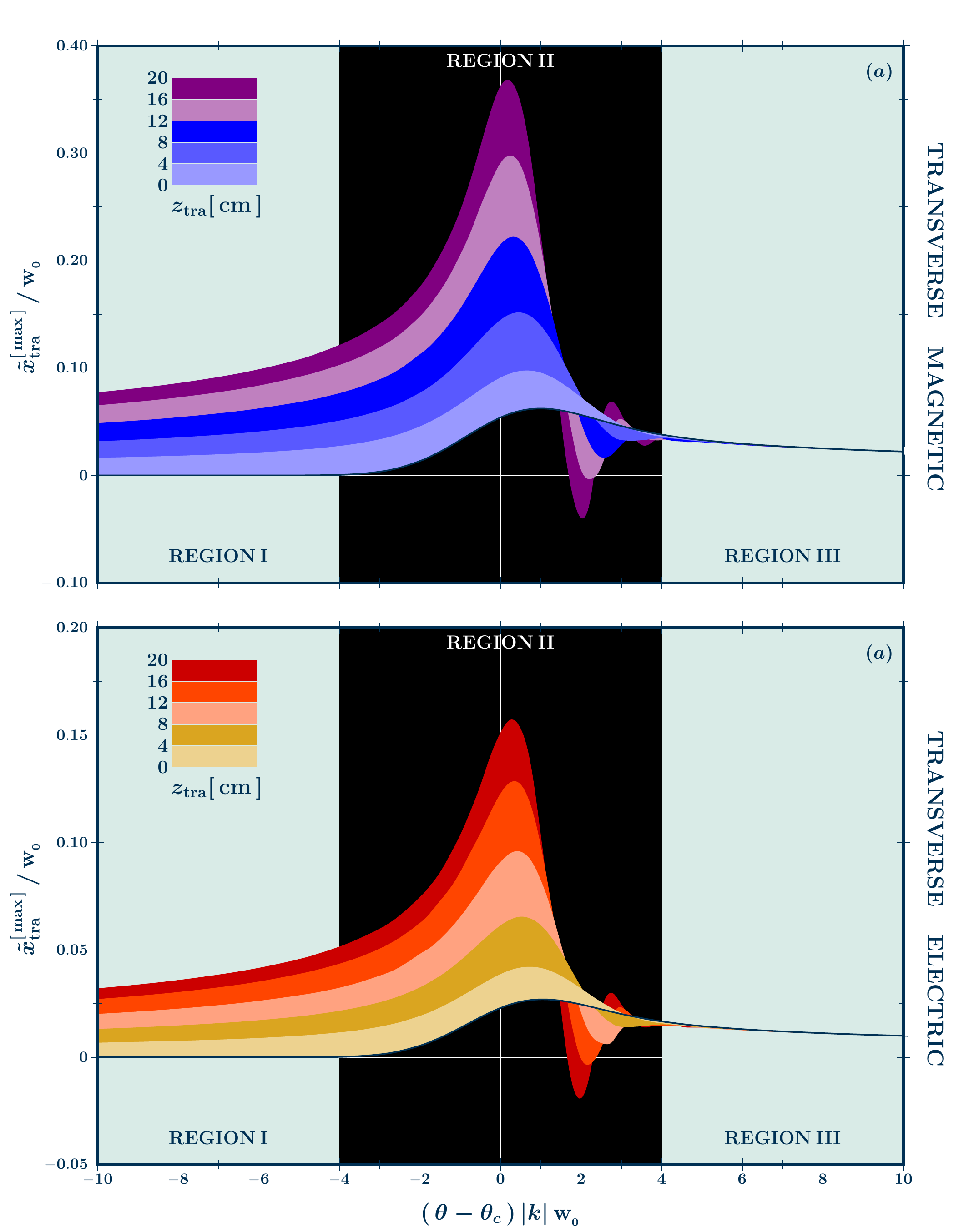}{{\small \textbf{Critical incidence region.}} Numerical lateral displacements  of the maximum of the upper transmitted beam plotted as function of the incidence angle,  
$\delta=\left(\,\theta-\theta_c\right)\,\mk\,\wo$ for different axial positions both for magnetic (a) and electric (b) waves.
The angular deviations and GH shifts refer to an optical Gaussian beam with 
$\wo=100\,\mathrm{\mu m}$, $\lambda=532 \,\mathrm{nm}$ and the  dielectric block to a BK7 prism,  $n\,=\,1.5195$. The black zone  represents the critical incidence region in which our analytical approximation fails due to the presence of an infinity in the Taylor expansion. In the incidence region I, $\delta<4$, and III ,$\delta>4$,   our analytical approximations show an excellent agreement with the numerical calculations. 
In region I, it is clear the axial dependence of the displacement  caused by angular deviations and in region III the lateral  displacement due to the GH shift. In region III, we do not see any angular deviations because the dominant contribution comes from the Fresnel coefficients of the internal reflection.  
 \label{fig2}}}

\def\figurethree{
\WideFigureSideCaption{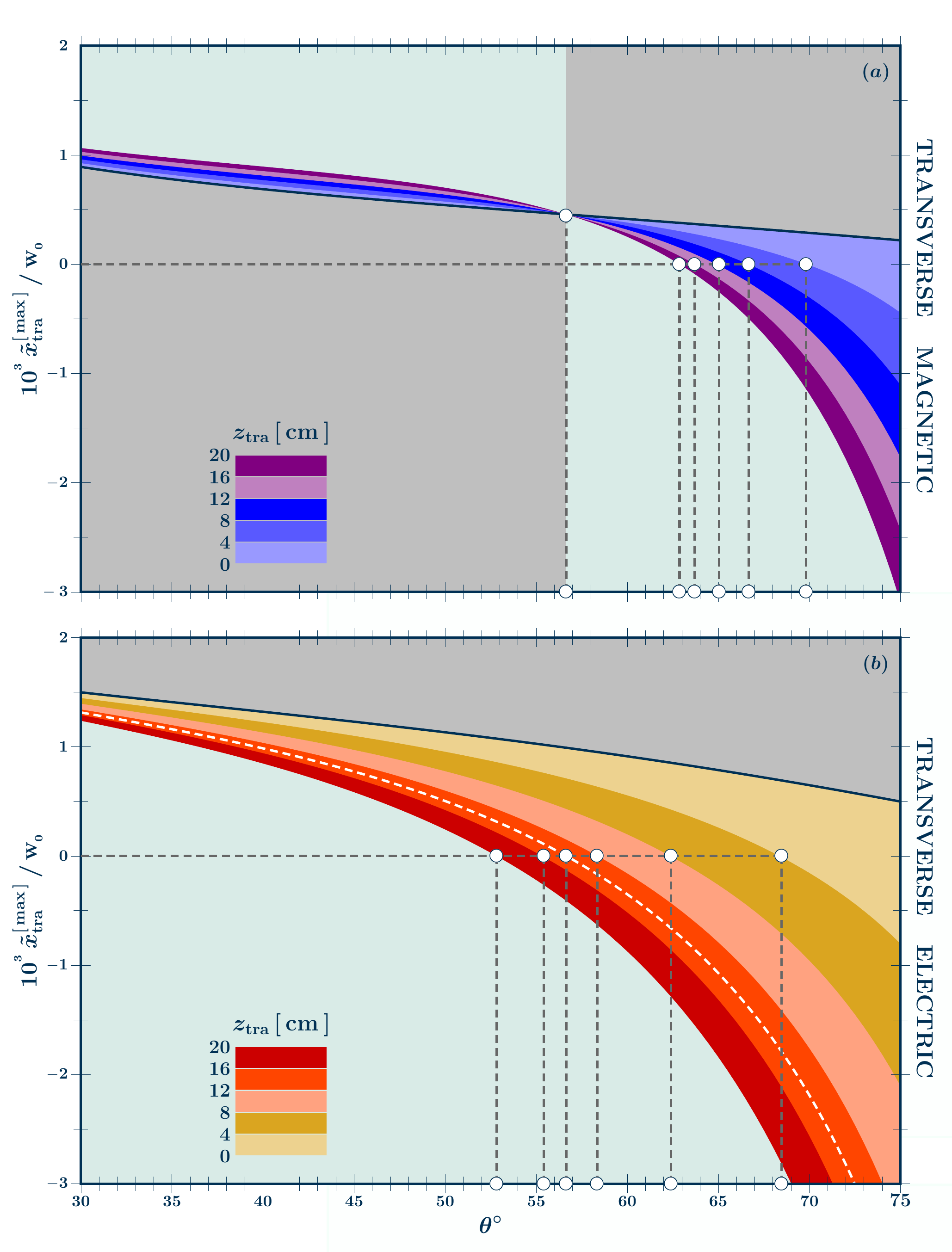}{{\small \textbf{GH shift and angular deviations as function of the incidence angle for different axial distances.}}
The displacement of the maximum of the upper transmitted beam is plotted for transverse magnetic and  electric waves, respectively in (a) and (b).  In (a), at Brewster incidence, $\theta_{_{\mathrm{B}}}=56.65^{o}$, the axial dependence is removed.  In (b), for Brewster incidence angular deviations compensate the GH shift at an axial distance of $14.14$ cm, white dashed line. The coloured zones refer to different axial distances. In (a) and (b), we also find the incidence angle for which
these optical effects offset each other.
\label{fig3}}}

\def\figurefour{
\WideFigureSideCaption{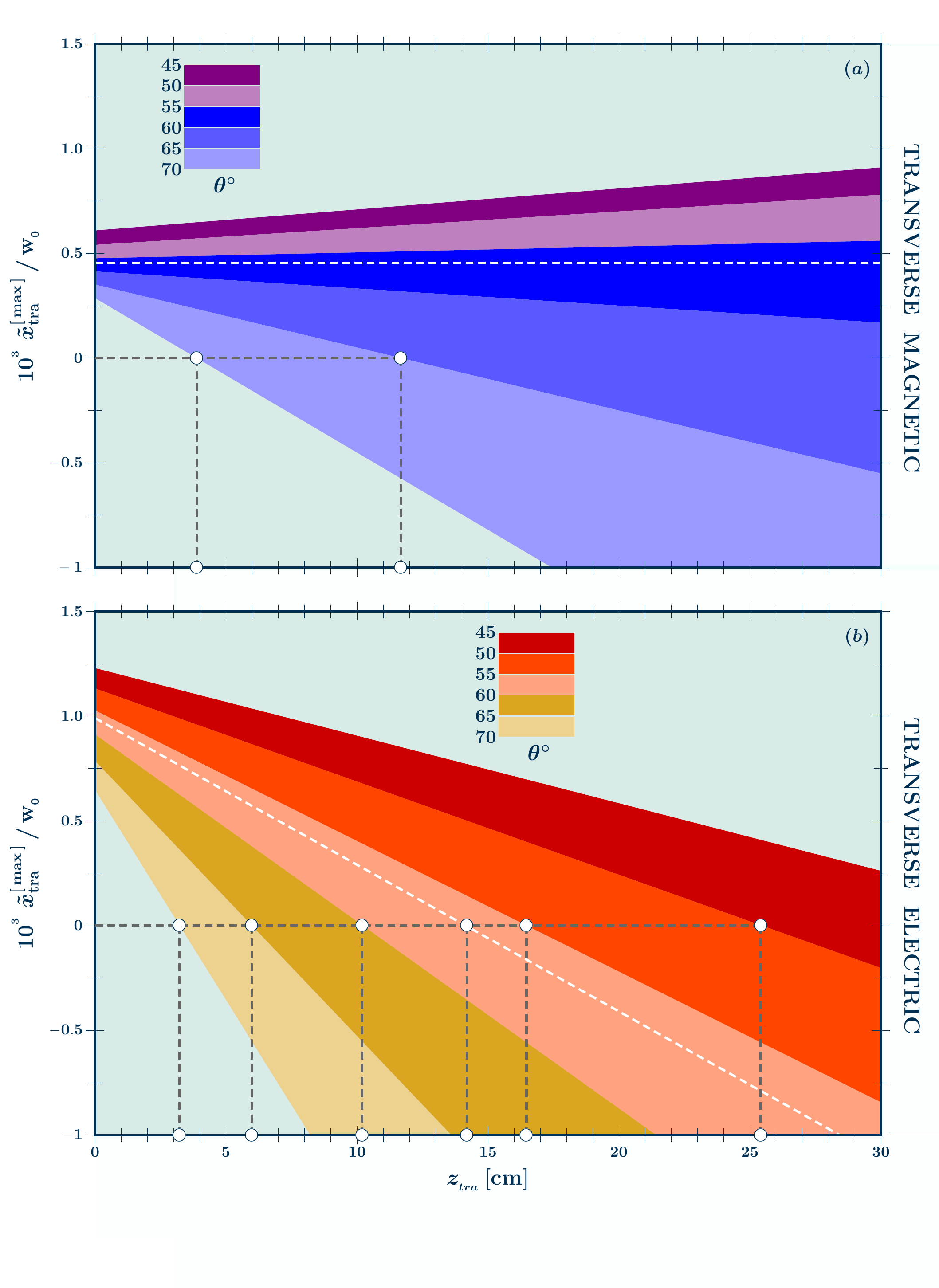}{{\small \textbf{GH shift and angular deviations as function of the axial distance for different incidence angles.}}
In (a) and (b), the white dashed line refer to the Brewster incidence. For transverse magnetic waves no axial dependence is seen. For electric waves, angular deviations compensate the GH shift at an axial distance of $14.14$ cm. The coloured zones refer to different incidence angles.  In (a) and (b), we also find the axial distance for which
these optical effects offset each other. \label{fig4}}}

\logo{
\colorbox{Teal}{\color{white}$\mathbf{\Sigma\hspace*{0.06cm} \delta \hspace*{0.04cm} \Lambda}$}
}

\journal{{\textbf{\color{DarkRed} Photonics} \, \textbf{{\color{PrussianBlue} 9}\,,\, 643-14 
\,\,({\color{PrussianBlue} 2022})\,.}}}

\titlelines{3}
\title{Goos-H\"anchen  lateral displacements and\\ angular deviations: When these optical\\ effects offset each other }

\imgbgabstract{For optical beams, transmitted by a right angle prism,  the Goos-H\"anchen shift can never be seen as a pure effect. Indeed, the lateral displacement, caused by the total internal reflection, will always be accompanied by angular deviations generated by the transmission through the incoming and  outgoing interfaces. This combined effect can be analysed by using the Taylor expansion of the Fresnel coefficients.  The analytic expression found for  the  transmitted beam allows to determine the beam parameters, the incidence angles, and the axial distance  for which lateral displacements are compensated  by angular deviations. Proposals to optimize  experimental implementations are also briefly discussed.
}

\author{
\names{Stefano De Leo\textsuperscript{1,3}, Luca Maggio\textsuperscript{2}, Moreno d'Ambrosio\textsuperscript{2}}
\affiliation{\textsuperscript{1}Department of Applied Mathematics, State University of Campinas, Brazil\\
\textsuperscript{2}Department of Mathematics and Physics, University of Salento, Italy
}
\email{\textsuperscript{3}deleo@unicamp.br}
}

\begin{document}

\sdlmaketitle

\section{Introduction}
%\subsection{Subsection test}

\figureone

\figuretwo

\figurethree

\figurefour

The interaction between optical beams and dielectric blocks has always been the subject of great interest, leading, in the past, to formulate the well-known laws of geometric optics\cite{born,sharma,saleh}. In the last century, new phenomena  like Goos-H\"anchen shift \cite{goos, artmann, crit1, crit2,crit3} and angular deviations \cite{ra} showed that the optical path predicted by geometric optics only represents an approximation to the real one. Theoretical studies  have been undertaken in order to understand in which situations lateral displacements and angular deviations can be amplified and then observed in the laboratory. The omnipresence 
of these phenomena\cite{metallic, waveguides, beammodes, sismic1, sismic2} also stimulated their application in technology \cite{micros, highsens, vapor}.

%Starting from the first experiment of Goos and Hänchen \cite{goos}, in 1947, phisicists began to investigate the deviation from the optical path of light. In particular,
In 1947, Goos and H\"anchen \cite{goos} were the first researchers to experimentally observe the lateral displacement of optical  beams transmitted, after many internal reflections, by a dielectric block. The experimental result, today known  as Goos-Hänchen shift, was, one year later, explained by Artman\cite{artmann}. Artmann’s observation was that  multiple plane waves, contributing to the final electromagnetic field, have rapidly varying phases that cancel each other out. Total internal reflection is indeed characterized  by a complex  Fresnel coefficient.  The stationary condition gives the main term of the phase which is responsible  for the additional  phase generating  the  lateral shift in the optical path \cite{s2015}. The  divergence in the Artmann formula was later removed \cite{crit1,crit2}. Recently, for incidence in the critical region,  an analytical formula, based on the modified Bessel functions, was proposed in \cite{crit3} and, some years later,  experimentally confirmed \cite{confirmation}.

In 1973, Ra, Bertoni and Felsen  \cite{ra} introduced the phenomenon of angular deviation. This phenomenon appears both for transmission (in this case, we have deviations from the refraction angle predicted by the Snell law) and partial reflection (in this case, we find deviations from the reflected angle predicted by the reflection law).  This phenomenon  is due, essentially, to the \textit{symmetry breaking} of the Gaussian distribution caused by the Fresnel coefficients modulating the Gaussian distribution  in the integral form of the transmitted and reflected beams.  

Angular deviations and Goos-H\"anchen sfhits  have been investigated in great detail in different fields, not only in optics \cite{metallic,waveguides,beammodes,micros,vapor} but also in seismic data analysis\cite{sismic1, sismic2}. In the critical region, lateral displacements and angular deviations generate oscillatory phenomena,  theoretically predicted in \cite{oscill1} and, recently, experimentally confirmed in \cite{oscill2, oscill3}.

In this article, we analyse the combined effect of the angular deviations (caused by the  transmission through  the incoming and outgoing triangular prism interfaces) and  the Goos-H\"anchen shift (caused by the total internal  reflection). The study is done outside the critical region. This choice is justified because, outside the critical region, we have  the possibility  to find  an analytic expression for the transmitted beam by using the Taylor expansion of the Fresnel coefficients and, consequently,  determine the beam parameters, the incidence angles, and the axial distance for which angular deviations compensate Goos-H\"anchen lateral displacements.  The integral form of the  beam transmitted through a dielectric prism,  see Fig.\,\ref{fig1}(a), is characterised  by three Fresnel coefficients:  the ones corresponding to the  transmission at the left (air/dielectric) and right (dielectric/air) interfaces and the one corresponding to the total internal reflection at the lower (dielectric/air) interface. The upper transmitted beam is, thus, the  perfect candidate to study the combined effect of angular deviations and Goos-H\"anchen shifts. In the next section, we fix our notation, introduce the Fresnel coefficients, and  calculate the phase of the optical beams. The integral form of the (upper) transmitted beam cannot be analytically solved, so we use the Taylor expansion of the Fresnel coefficients and of the optical phase  to obtain a closed form for the transmitted beam. By using this analytic approximation, we obtain a \textit{cubic equation} which allows to determine the peak position  of the transmitted beam. {\color{Red}{In a previous paper\cite{alessia}, based on this cubic equation, we studied the phenomenon of  pure angular deviations, this implies incidence angle below the critical one. In this paper we analyse incidence greater than the critical one. This allows to investigating both angular deviations and Goss-H\"anchen displacements (only present in the case of total internal reflection). In this incidence region, it is thus possible to study when these optical effect offset each other.}}   
Discussions,  conclusions, and proposals for experimental implementations appear in the final sections. 

\section{The incident beam}

Let us introduce the  integral form of the incident beam

\begin{equation}
\label{incB}
E\in(\mathbf{r})\, =\, E_{\0}\int\hspace*{-.1cm} \d \kx \,\d \ky \,\,G(\kx,\ky) \, \,
e^{{\,i \,\boldsymbol{k}\,\cdot\, \mathbf{r}}}\,\,\,,
\end{equation}
where
\begin{equation*}
G\left(\kx,\ky\right) \,=\, \frac{\wo^{\2}}{4\,\pi}\, \exp \left[\,-
\,\left(\,\kx\q +\ky^{^{2}}\,\right)\,\frac{\wo^{\2}}{4}\,\right]
\end{equation*}
is the Gaussian wave number distribution, and
\[\boldsymbol{k}\,\cdot\, \mathbf{r} \,=\, \kx x + \ky y + \kz z\]
is the optical phase with $\mk = 2\,\pi/\lambda$. By using the paraxial approximation,  
 \[ k_z\,\approx\, \mk -  (\,\kx^{^{2}}+\ky\q\,)\,/\,2\,\mk\,\,,\] 
the  integral in Eq.\,(\ref{incB})   can analytically be solved leading to the following closed expression for
the incident  Gaussian beam
\begin{equation}
E\in(\mathbf{r})\,=\, \frac{E_{\0}\,e^{{\,i\,\mk\,z}}}{1 + i\,z/\zR}\,\exp \left[\,-\,\,\frac{x\q+y^{^{2}}}{\wos\, (1 + i\,z / \zR)}\,\right]\,\,\,,
\end{equation}
where $\zR = \pi\wo\q/\lambda$ is the Rayleigh axial range {\color{Black}{and $\wo$ the beam waist}}. The beam intensity is then given by
\begin{equation}
I\in(\mathbf{r})  \,=\, I_{\0} \,\frac{\wo\q}{\w\q(z)} \, \exp \left[\,- \,2 \,\,\frac{ x\q + y\q}{\w\q(z)}\,\right]\,\,\,,
\end{equation}
where $I_{\0}=E_{\0}\,\q$ and
$\w(z) = \wo \, \sqrt{1 + (z/\zR)\q}$.

\section{The optical phase}

In the integral form of optical beams, an important role is played by the optical phase responsible for the optical path of the beam.  In order to calculate the optical phase of the (upper) transmitted beam, it is useful to introduce the coordinate system corresponding to the incident and transmitted beams and the ones corresponding to the left, right, and lower  interfaces, see Fig.\ref{fig1}(b), 
\begin{equation*}
\label{rotcord}
\begin{pmatrix}
\xt \\ 
\zt \\
\end{pmatrix}
= 
M\left(\,-\theta\,\right)
\begin{pmatrix}
x \\
z \\
\end{pmatrix}
=
M\left(\,\frac{\pi}{4}\,\right)
\begin{pmatrix}
\xs \\
\zs \\
\end{pmatrix}
=
M\left(\,\theta\,\right)
\begin{pmatrix}
\ztr\, \\
\xtr \\
\end{pmatrix}\,\,,
\end{equation*}
where 
{\color{Black}{$M(\theta)=\,\{\,\{\,\cos\theta\,,\,-\,\sin\theta\,\}\,,\,\{\,\sin\theta\,,\,\cos\theta\,\}\,\}$}} 
represents the anti-clockwise rotation matrix. The optical phase corresponding to the beam propagating from the source to the first interface is given by
\begin{center}
\begin{tabular}{l l}
\circledO{S}{\CS} $\,\,\rightarrow\,\,$ \circledM{1}{C1} \,\,  :       & $\kx\, x\,+\,\kz\, z\,=\,\kxt\,\xt\,+\,\kzt\,\zt$ \,\,,
\end{tabular}
\end{center}
where
\[\begin{pmatrix}
\kxt \\ 
\kzt \\
\end{pmatrix}
\, = \, 
M\,\left(\,-\theta\,\right)
\begin{pmatrix}
\kx \\
\kz \\
\end{pmatrix}\,\,\,.
\]
After transmission through the left (air/dielectric) interface, the beam moves, into the dielectric, towards the lower (dielectric/air) interface with  the following optical phase 
\begin{center}
\begin{tabular}{l l}
\circledM{1}{C1} $\,\,\rightarrow\,\,$ \circledM{2}{C2}   \,\,:       & $\qxt\,\xt\,+\,\qzt\,\zt\,=\,\qxs\,\xs\,+\,\qzs\,\zs$\,\,,
\end{tabular}
\end{center}
where
\begin{equation*}
\begin{pmatrix}
\qxt \\
\qzt \\
\end{pmatrix}
\,=\,
M\,\left(\,\frac{\pi}{4}\,\right)
\begin{pmatrix}
\qxs \\
\qzs \\
\end{pmatrix}\,\,,
\end{equation*}
with
\[\qxt\,=\,\kxt\,\,\,\,\,\mathrm{and}\,\,\,\,\,\qzt\,=\,\sqrt{\displaystyle n\q\,\mk\q-\qxt\q-\ky\q}\,\,.\]
The beam is then reflected back and  moves between the lower and right interface with an optical phase  given by 
\begin{center}
\begin{tabular}{l l}
\circledM{2}{C2} $\,\,\rightarrow\,\,$ \circledM{3}{C3} \,\,  :       & $\qxs\,\xs\,-\,\qzs\,\zs\,=\,\qxt\,\zt\,+\,\qzt\,\xt$\,\,.
\end{tabular}
\end{center}
Finally, in the  integral form of the (upper) transmitted beam appears, as expected, the following optical phase  
\begin{center}
\begin{tabular}{l l}
\circledM{3}{C3} $\,\,\rightarrow\,\,$ \circledO{C}{\CC}\,\,   :       & $\kxt\,\zt\,+\,\kzt\,\xt\,=\,\kx\,\xtr\,+\,\kz\,\ztr$\,\,.
\end{tabular}
\end{center}

\section{The upper transmitted beam}

Once obtained the optical phase of the upper transmitted beam, we can write its integral form:
\begin{align}
E\pol\tr(\rtr)\, = \,E_{\0}\int\hspace*{-.1cm} \d \kx \,\d \ky \,\,\Gtra \, \,
e^{{\,i \,\boldsymbol{k}\,\cdot\, \rtr}}\,\,,
\label{Etra0}
\end{align}
where $\,\,\rtr=(\,\xtr,y,\ztr\,)$ and   
\[\Gtra \,=\,T\pol(\kx,\ky)\,G(\kx,\ky)\,\,,\] 
with
\begin{eqnarray*}
T\pol(\kx,\ky)\, =\,  \frac{4\,\kzt\qzt}{(a\pol\kzt+\qzt/a\pol)\q}\,
\frac{\qzs/a\pol-a\pol\kzt  }{\qzs/a\pol+a\pol\kzs }\,\,\times \nonumber\\
  \exp\{\,i\,[\, \qzs d\,\sqrt{2}\,+\,(\,\qzt\,-\,\kzt\,)\,(\,l\,-\,d\,)\,]\,\}\,\,,
 \end{eqnarray*}
($a\tm = n$ and $a\te = 1$). The additional phase appearing in the Fresnel coefficients is due to the fact that the discontinuities at the air/dielectric and dielectric/air interfaces are located at different points. This phase is responsible for the optical path predicted by geometric optics.

In order to integrate Eq.(\ref{Etra0}), we use the first order Taylor expansion of the transmission coefficient, i.e.
\begin{eqnarray}
T\pol(\kx,\kx)&=& T\pol(0,0)\, \left[\,1\,+\,\beta\pol \,\frac{\kx}{\mk}\,\right]\,\times \nonumber \\
& & \exp[\,-\,i\,\kx\,x\sn\,]\,\,,
\label{taylor}
\end{eqnarray} 
where
\begin{eqnarray*}
T\pol(0,0)&=&\frac{4\,n\cos\theta\,\cos\psi}{(a\pol\cos\theta+n\cos\psi/a\pol)\q}\,\times\\
 & & 
\frac{n\cos\varphi/a\pol-a\pol\cos\phi}{n\cos\varphi/a\pol+a\pol\cos\phi}\,\times\\
&&\hspace*{-2cm}\exp\{\,i\,[\, n \,\cos\varphi \,d\,\sqrt{2}\,+\,(\,n\,\cos\psi\,-\,\cos\theta\,)
 (\,l\,-\,d\,)\,\mk\,]\,\} 
\end{eqnarray*} 
 and
\[x\sn=(\,\tan\psi\cos\theta\,-\,\sin\theta\,)\,l\,+\,(\,\cos\theta\,+\,\sin\theta\,)\,d\,\,.\]
The $\beta\pol$ factor in (\ref{taylor}) can be expressed  in terms of 3 addends, respectively, 
corresponding to the transmission through the  left (air/dielectric) interface, $\circledM{1}{C1}$, to the reflection by the lower (dielectric/air) interface,  $\circledM{2}{C2}$, and, finally, to the transmission through the right (dielectric/air) interface,  $\circledM{3}{C3}$, 
\[\beta\pol=\beta\pol\ua+\beta\pol\ub+\beta\pol\uc\,\,,\]
with
\begin{eqnarray*}
\beta\te\ua&=&\tan\psi\,-\,\tan\theta\,\,,\\
\beta\te\ub&=& 2\,\tan\phi\,\,\varphi'\,\,,\\
\beta\te\uc&=&(\,\tan\theta\,-\,\tan\psi\,)\,\,\psi'\,\,,\\
\beta\tm\ua&=&(\,\tan\psi\,-\,\tan\theta/\,n\qq\,)\,/\,(\,\sin\qq\psi\,-\,\cos\qq\theta\,)\,\,,\\
\beta\tm\ub&=&2\,\tan\phi\,\,\varphi'\,/\,(\,\sin\qq\phi\,-\,\cos\qq\varphi\,)\,\,,\\
\beta\tm\uc&=&(\,\tan\theta\,-\,n\qq\tan\psi\,)\,\,\psi'\,/\,(\,\sin\qq\theta\,-\,\cos\qq\psi\,)\,\,,
\end{eqnarray*} 
where the different angles which appear in the previous formulas are related to the incidence angle $\theta$ by the Snell law, i.e. $\sin\theta=n\,\sin\psi$ and  $n\,\sin\varphi=\sin\phi$, the angle $\varphi$ to 
$\psi$ by the geometry of the prism, i.e. $\varphi=\psi+\pi/4$. Finally, we have $\varphi'=\psi'=\cos\theta/n\,\cos\psi$.
 
By using the Taylor expansion (\ref{taylor}), we can analytically solve the integral of Eq.\,(\ref{Etra0}).
The $k_x$ term in the exponential will be responsible for the  shift in the $\xtr$ coordinate, i.e. 
\[\,\xttr\,=\,\xtr\,-\,x\sn\,\,,\]
centring the Gaussian beam in the optical path predicted by the Snell and reflection laws. 
{\color{Black}{The constant term in (\ref{taylor}), i.e. $T\pol ( 0 , 0 )$, leads to the same integration done
for the incident, consequently we obtain the following contribution
\[
T\pol ( 0 , 0 )\,\,E\in \left(\,\mathbf{\rttr} \right)\,\,.
\]
The  linear term, i.e. $T\pol ( 0 , 0 )\,\beta\pol \,\kx\,/\,\mk$, is responsible for the breaking of the Gaussian symmetry 
for incidence below the critical one and for  the Goos-H\"anchen shift in the case of total internal reflection.  Observing that $k_x$ in the integrand of (\ref{Etra0}) can be replaced by $-\,i\,\partial/\partial \xttr$, we obtain the following contribution
 \[
-\,i\,T\pol ( 0 , 0 )\,\,\frac{\beta\pol}{\mk}\,\,\frac{\partial\,E\in \left(\,\mathbf{\rttr} \right)}{\partial \xttr}\,\,.
\]
The analytical expression, for the upper transmitted beam, is  then given by
\begin{eqnarray*}
E\tr\pol(\,\mathbf{\rttr}\,) & = &  \left[\,1 +\,2\,i\,\,\frac{ \beta\pol \xttr}{\mk\,\wo^{^{2}}\,(\,1 + i\, \ztr / \zR\,)}\, \right]\,\times \nonumber \\
 & & T\pol ( 0 , 0 )\,\, E\in \left(\,\mathbf{\rttr} \right)\,\,.
\end{eqnarray*}
}}
Finally, after algebraic manipulations, we find
\begin{eqnarray}
\label{Etra}
E\tr\pol(\,\mathbf{\rttr}\,) 
&= &   \left(\,1 +\,i\,\, \frac{ \beta\pol \xttr +\ztr}{\zR}\, \right)\,\times \nonumber \\
& & \frac{T\pol ( 0 , 0 ) }{1 + i\, \ztr / \zR}\,\, E\in \left(\,\mathbf{\rttr} \right)\,\,.
\end{eqnarray}
In order to check the validity of our analytical approximation, let us briefly analyse what happens near the critical incidence region. The critical angle is found when $n\,\sin\varphi_c=1$, this implies a critical incidence at 
\begin{equation}
\theta_c\,=\,\arcsin\left[\,\left(\,1\,-\,\sqrt{n\q-1}\,\right)\,/\, \sqrt{2}\,\,\right]\,\,.
\end{equation}
In Fig.\,2, we plot the the (upper) transmitted beam shift of the maxima with respect to the path predicted by geometric optics. This is done by numerically integrating  Eq.\,(\ref{Etra0}). The plots of the maxima, as a function  of 
$\delta=\left(\theta-\theta_c\right)\mk\wo$,  refer to a Gaussian laser with  $\wo\,=\,100 \,\mathrm{\mu m}$,  $\lambda\,=\,532 \,\mathrm{nm}$ and $n\,=\,1.5195$ (BK7 prism).  We can distinguish three regions. 

Region I, before the critical region, shows an axial dependence of the shift and this is caused by the modulation of the Gaussian wave number function  generated by the real Fresnel coefficients related to the transmission through the first and third interface and the {\color{Black}{partial}} internal reflection. These phenomena represent angular deviations to the Snell and reflection law of geometric optics, 
{\color{Black}{For a detailed discussion of pure  angular deviations and its amplifications near the Brewster incidence, we refer the reader to the article cited in \cite{alessia}}}.

Region II determines  the critical region, in such a region   
the infinity in  $\beta\ub\pol$ coefficients required a more complicated technique of integration to obtain the analytical expression for the upper transmitted beam \cite{crit3} and new oscillatory phenomena appear \cite{oscill1,oscill3}.  

In region III, for incidence greater than the  critical one  but near enough to amplify the GH shift with respect to angular deviations, this axial depends breaks down.  {\color{Black}{Region III will be the region of interest for our discussion because in this region, far enough of the critical region, angular deviations and GH shifts can offset each other. The analysis in this region complements the one presented in ref.\,\cite{alessia}}}.   In region III,  we have
\[\tan\phi = n\,\sin\varphi\,/\,i\,\sqrt{n^{\2}\sin^{\2}\varphi-1}\,\,,\]
and consequently the intensity of the upper transmitted beam can be written in the following form

\begin{align}
\label{transmittedintensity}
I\tr\pol\,(\mathbf{\rttr})\,=\,\frac{\wo\q}{\mathrm{w}\q\,(\ztr)}\,\,T\q\pol\,(0\,,0)\,I\in (\mathbf{\rttr})\,\times\nonumber
\\
\left[\left(1+\frac{\gamma\ub\pol\,\xttr}{\zR}\right)\q+\left(\frac{\ztr+\beta\uac\pol\,\xttr}{\zR}\right)\q\,\right]\,\,,
\end{align}
where
\begin{eqnarray*}
\gamma\te\ub&=& 2\,\,\frac{n\,\sin\varphi}{\sqrt{n^{\2}\sin^{\2}\varphi-1}}\,\,\cos\theta\,/\,n\,\cos\psi\,\,,\\
\gamma\tm\ub&=&2\,\,\frac{n\,\sin\varphi}{\sqrt{n^{\2}\sin^{\2}\varphi-1}}\,\,\frac{\cos\theta\,/\,n\,\cos\psi}{n\q\sin\q\varphi\,-\,\cos\q{\varphi}}\,\,.
\end{eqnarray*} 
and
\begin{equation*}
\beta\uac\pol=\,\beta\ua\pol+\beta\uc\pol\,\,.
\end{equation*}

\section{GH shifts and angular deviations}

The analytical expression found for the intensity of the upper transmitted beam, see  Eq.\, \eqref{transmittedintensity}, allows to calculate its maximum and consequently to
obtain the lateral displacement with respect to the path predicted by geometric optics due to the  GH shifts and angular deviations. The  intensity $\xttr$ derivative   leads to the following cubic equation
\begin{equation}
\label{cubic}
{\left(\frac{\xttr}{\wo}\right)}^{^3}\,+\,a\pol\,{\left(\frac{\xttr}{\wo}\right)}^{^2}\,+\,b\pol\,\frac{\xttr}{\wo}\,=\,c\pol\,\,,
\end{equation}
where
\begin{eqnarray*}
a\pol&=&2\,\, \dfrac{\gamma\ub\pol\,\zR\,+\,\beta\uac\pol\,\ztr}{\left({\gamma\ub\pol}\q+{\beta\uac\pol}\q\right)\wo}\,\,,
\\
\\
b\pol&=&\frac{\w\q (z)}{\wo\q}\,\left[ \dfrac{\zR\q}{\left({\gamma\ub\pol}\q+{\beta\uac\pol}\q\right)\wo\q}-\frac{1}{2}\right]\,,
\\
\\
c\pol&=&\frac{\w\q (z)}{\wo\q}\,\dfrac{\gamma\ub\pol\,\zR\,+\,\beta\uac\pol\,\ztr}{2\,
\left({\gamma\ub\pol}\q+{\beta\uac\pol}\q\right)\wo}\,\,.
\end{eqnarray*}
This equation allows to calculate and compare the lateral displacements in region III. When the GH shifts dominate no axial dependence can be seen. When the angular deviations become comparable with GH shifts an axial dependence is seen in  the lateral displacements. 

Eq.\,(\ref{cubic}) can be reduced to a linear equation by observing that $\xttr\ll \wo$ and that 
{\color{Black}{$b\pol\gg a\pol$}}  for axial distance $\ztr\ll \zR^{^{2}}\,/\,\wo$. The lateral displacement of the maximum is then given by 
\begin{eqnarray}
\label{linear}
\xttr\max & = & c\pol\,\wo\,/\,b\pol \nonumber\\
 & \approx & \dfrac{\gamma\ub\pol\,+\,\beta\uac\pol\,\ztr\,/\,\zR}{\mk}\,\,,
\end{eqnarray}
where the axial independent term, proportional to $\lambda$, represents the pure  GH shift \cite{goos,artmann} and the axial dependent the angular deviations  due to the Fresnel transmission modulation of the Gaussian wave number distribution.

Near the critical region, 
\[\theta \,=\, \theta_c\,+\,\delta\,/\,\mk \wo\:\:\:\:\:\:\:\:\:[\,\delta \geqslant 4\,],\]
we have
\[n^{\2}\sin^{\2}\varphi\,-\,1\,\approx\, 2\, n\, \cos\varphi_c\,\varphi_c'\,\,\delta\,/\,\mk \wo\,\,.\]
In the example analysed in this paper, i.e. $\lambda=532\,\mathrm{nm}$ and $\wo\,=\,100\,\mu\mathrm{m}$,
$\delta\geqslant 4$ implies and incidence angle greater than  $\theta_c\,+\,0.2^{^{\mathrm{o}}}$. 
Observing that
\[\beta\uac\pol\,\ll\,\gamma\ub\pol\,\propto\,\sqrt{\mk\,\wo}\,\,, \]
and using the approximated expression for the $\gamma$ factors,   we obtain
\begin{equation}
\label{amp}
\xttr\max\,=\,\frac{\sigma\pol}{n}\,\,\sqrt{\frac{2\,\cos\theta_c}{\delta\,\cos\varphi_c\,\cos\psi_c}\,\,\frac{\wo}{\mk}}\,\,,
\end{equation}
with $\{\,\sigma\te\,,\,\sigma\tm\,\}\,=\,\{\,1\,,\,n^{\2}\,\}$.
Clearly the axial dependence has been removed and this agrees with the numerical calculation shown in Fig.\,2, see region III at the right of the black zone. In  this region, Eq.\,(\ref{amp}) also contains the  well known $\sqrt{\mk\,\wo}$ amplification for the GH shift, for details see refs.\cite{crit3,s2013}.  The $\sigma$ factor is, finally, responsible  for a further amplification of $n^{\2}$  for the transverse magnetic wave, see the scale in Fig.\,2 (a) and (b).

For transverse magnetic waves, the pure Goos-H\"anchen shift is found for incidence at the Brewster angle, i.e.
\[\beta\uac\tm=0\,\,\,\Rightarrow\,\,\,\theta=\theta_{_{\mathrm{B}}}=\arctan n\,\,,\]
see Fig.\,3(a). For a given axial distance, from Eq.\,(\ref{linear}), we can obtain the incidence angle  for which the GH shift is compensated by  the angular deviation. For example, for a camera positioned at an axial distance of 
\[4,\,8,\,12,\,16,\,20\,\,\,\mathrm{cm}\,\,,\]
for the optical beam considered in this paper,   we find incidence angles  of 
 \[69.86^{o},\,66.65^{o},\,64.89^{o},\,63.72^{o},\,62.87^{o}\] 
 for transverse magnetic waves, see Fig.\,3(a), and of   
 \[68.36^{o},\,62.37^{o},\,58.37^{o},\,55.32^{o},\,52.82^{o}\] 
 for transverse electric waves, see Fig.\,3(b). Eq.\,(\ref{linear}) can be also used to find, for a given incidence angle, the axial distance for which 
GH lateral displacements and angular deviations offset each other,
\begin{equation}
\label{ztra}
\ztr\,\,=\,\,-\, \frac{\gamma\ub\pol}{\beta\uac\pol}\,\,\zR\,\,. 
\end{equation}
For example, for incidence angles of 
\[45^{o},\,50^{o},\,55^{o},\,60^{o},\,65^{o},\,70^{o}\,\,,\] 
the compensation happens, for transverse electric waves, at the axial distances
\[38.10,\,25.45,\,16.47,\,10.22,\,5.99,\,3.23\,\,\,\mathrm{cm}\,\,,\]
see Fig.\,4(b). For transverse magnetic waves, the compensation happens for incidence angles greater than the Brewster angle, $\theta_{_{\mathrm{B}}}=56.65^{o}$. For incidence angle of 
 \[60^{o},\,65^{o},\,70^{o}\,\,,\] 
angular deviations compensate the GH shifts at the axial distances
\[50.68,\,11.68,\,3.88\,\,\,\mathrm{cm}\,\,,\]
 see Fig.\,4(a).

\section{Discussions}

Lateral displacements of optical beams with respect to the path predicted by geometric optics stimulated, in the last decades, both theoretical and experimental investigations. Two types of displacements characterize the transmission through dielectric blocks. The first, known as GH shift, is due to the phase of the total internal reflection coefficient  
and  it is independent of the axial position of the detector. The second one is due to the modulation of
the transmission coefficients on the wave number distribution of the incident beam  and it is dependent on the axial position of the detector.  

In region III, far enough to  the critical region II, GH shifts are proportional  to the wavelength of the optical beam.
{\color{Black}{When the axial distance approaches the   Rayleigh axial range also the angular deviations become proportional to the wavelength and this open the doors to the possibility to cancel the lateral displacements induced by the total reflection coefficient. This phenomenon is also known as composite GH effect \cite{WM1,referee1}. In region I, where the  partial internal reflection implies the only presence of  angular deviations\cite{referee2} an amplification effect happens near the internal Brewster angle, for details see  ref.\, \cite{alessia}. Region II represents the region around the critical angle and an  amplification by a factor $\sqrt{\mk\,\wo}$, see Eq.\,(\ref{amp}), is found  in proximity of the critical incidence \cite{crit3,s2013}.  Such a region is also characterized by oscillatory phenomena \cite{oscill1,oscill2,oscill3} and the analytical analytical formula, obtained in this paper for the intensity of upper transmitted beam, i.e Eq.\,(\ref{transmittedintensity}), fails to reproduce the numerical data. It is important to observe here that region II represents a very  small region of the incidence spectrum covering a  range of $8/\mk\,\wo$ around the critical angle. This means, for a beam waist of $100$ $\mu$m and a wavelength of $532$ nm, 
a range of $0.4^{\mathrm{o}}$ around the critical angle. Consequently, the analytical formula presented in this paper is in excellent agreement with the numerical data for all the incidence angles greater than $\theta_c\,+\,4\,/\,\mk\,\wo$ or in the case of the beam parameters used in our simulations for incidence angles greater than   $\theta_c\,+\,0.4^{\mathrm{o}}$.
}}

\section{Conclusions and outlooks}

In this paper, by using the Taylor expansion of the Fresnel coefficients of the transmission through the first and third interfaces and of the total reflection by the second interface, we have given an analytical expression for the upper transmitted beam intensity, see Eq.\,(\ref{transmittedintensity}). From this analytical approximation it is  immediate to obtain the cubic equation to calculate the intensity maximum. The cubic equation (\ref{cubic}) can then be further reduced to a linear equation (\ref{linear}) from which we can obtain the incidence angles and axial distances for which GH shifts and angular deviations offset each other. For transverse magnetic waves this compensation effect is only possible for incidence greater than the Brewster incidence, $\theta_{_{\mathrm{B}}}=\arctan[n]$.

The analytical expression of the upper transmitted beam given in this paper, see Eq.\,(\ref{Etra}), is also useful in view of experimental implementations done by using the weak measurements technique \cite{WM1, referee3}. This technique is based on the interference between transverse electric and magnetic waves \cite{WM2,WM3}. Consequently the analytical expression for the upper transmitted beam is important to find the main maximum of the combined optical beam which is a function of the different lateral displacements and angular deviations of transverse electric and magnetic waves. For the incidence angles and axial distances for which these optical effects offset each other, the weak measurement breaks down because the double peak phenomenon is no longer present. In a forthcoming paper, we shall revise the weak measurements for transmission through dielectric blocks in view  of the analytical expression given in this article.

\subsection*{Acknowledgements}
One of the authors (S.D.L.) thanks the CNPq (grant 2021/307664) and Fapesp (grants
2019/06382-9 and 2021/08848-5) for the financial support and the University of Salento (Lecce)  
for the hospitality. The authors are also grateful to A. Alessandrelli, L. Solidoro and A. Stefano for their scientific comments and suggestions during the preparation of this article and to Profs. G. C\'o, L. Girlanda, M. Martino, and M. Mazzeo for their help in consolidating the research BRIT19 project, an  international collaboration between the State University of Campinas (Brazil) and the Salento University of Lecce (Italy).

\end{document}